\documentclass[aps,pre,twocolumn,showpacs,superscriptaddress,longbibliography]{revtex4-1}
\usepackage{amsmath,amssymb}
\usepackage{graphicx,color}
\usepackage{times}
\usepackage[pdftex,bookmarks,colorlinks,breaklinks]{hyperref}% PDF hyperlinks, with colored links
\makeatletter
\hypersetup{linkcolor=blue,citecolor=blue,filecolor=dullmagenta,urlcolor=blue, % colored links
 pdftitle={Statistics of a simple transmission mode on a lossy chaotic background},
 pdfauthor={Dmitry V. Savin}, pdflang={English} }

\newcommand{\aver}[1]{\left \langle #1 \right \rangle}

\newcommand{\Hbg}{H_{\mathrm{bg}}}
\newcommand{\Szero}{S^{(0)}_{aa}}
\makeatother
\begin{document}

\title{Statistics of a simple transmission mode on a lossy chaotic background}
\author{Dmitry V. Savin}
 \affiliation{Department of Mathematics, Brunel University London, Uxbridge, UB8 3PH, United Kingdom}
%\date{\today}
%\received{}
\published{3 March 2020 in \underline{{Phys. Rev. Research \textbf{2}, 013246 (2020)}}\ \ }

\begin{abstract}
Scattering on a resonance state coupled to a complicated background is a typical problem for mesoscopic quantum many-body systems as well as for wave propagation in the presence of a complex environment. On average, such a simple mode acquires an effective damping, the so-called ``spreading'' width, due to mixing with the background states. Modeling the latter by random matrix theory and employing the strength function formalism, we derive the joint distribution of the reflection and total transmission at arbitrary absorption in the background. The distribution is found to possess a remarkable symmetry between its reflection and transmission sectors, which is controlled by the ratio of the spreading to escape width. This in turn results in a symmetry relation between the marginal densities, despite the absence of the flux conservation at finite absorption. As an application, we study the statistics of total losses in the system at arbitrary coupling to the background.
\end{abstract}
%\pacs{05.45.Mt, 03.65.Nk, 05.60.Gg, 24.60.-k}
% 05.45.Mt Quantum chaos; semiclassical methods
% 03.65.Nk Scattering theory
% 05.60.Gg Quantum transport
% 24.60.-k Statistical theory and fluctuations
\maketitle

\section{Introduction} %
Strength function phenomena \cite{soko97} have a rich history of various applications in atomic and nuclear physics \cite{Bohr,harn86,soko97i,gu99,zele16} as well as in open mesoscopic systems \cite{aber08,soko10,mora12,savi17}. In such problems, one deals with a ``simple'' excitation (associated with a specific signal) that is coupled to the background of many ``complicated'' (usually chaotic) states. As a result of this coupling, the simple mode is spread over exact stationary states with a rate determined by the so-called spreading width \cite{soko97, Bohr}. Transmission through such a simple mode is therefore characterized by the competition between the two damping mechanisms, escape to the continuum and spreading over the background, and becomes strongly suppressed when the ratio $\eta=\Gamma_{\downarrow}/\Gamma_0$ of the spreading ($\Gamma_{\downarrow}$) to escape ($\Gamma_0$) width exceeds unity \cite{savi17}.

Under real laboratory conditions, there are also sources of a coherence loss in quantum transport, with finite absorption being one of them \cite{kuhl05a}. This has dramatic consequences in scattering, since the $S$ matrix becomes no longer unitary. For open quantum or wave chaotic systems, exact analytical results were recently obtained for various scattering characteristics at finite absorption \cite{savi03a,fyod03,fyod05,kuma13}. Recent advances in experimental techniques have made it possible to change absorption in a controlled way and to test the theory with high accuracy in microwave cavity experiments \cite{kuhl13}, including in particular the statistics of reflection and transmission coefficients~\cite{kuhl05,koeb10,diet10}, complex impedances \cite{hemm05,hemm06, grad14}, and decay rates~\cite{kuhl08}.

On the theory side, the resonance scattering formalism \cite{Mahaux} is well adopted to treat both dynamical and statistical features of such systems on equal footing \cite{verb85,soko89,fyod97}. When combined with random matrix theory (RMT) to model internal chaotic dynamics~\cite{Stoeckmann,guhr98}, it offers a powerful tool to describe universal fluctuations in scattering, see Refs.~\cite{mitc10,fyod11ox} for recent reviews. The approach is also flexible in incorporating system-specific effects. In particular, the simple mode in such a context was recently introduced as a useful model for quantifying fluctuations induced by complex environments in the transmission intensity \cite{savi17} and phase \cite{savi18}. For the complete characterisation of the scattering process, however, both transmission and reflection fluctuations need to be treated at the same time. This becomes even more challenging at finite losses, since the two are no longer related by the flux conservation.

Here, we develop a general approach to scattering on the simple mode coupled to a lossy chaotic background. We derive exact results for the joint distribution of reflection and total transmission at arbitrary absorption. The distribution is shown to have a specific symmetry between its reflection and transmission sectors under the involution $\eta\to\eta^{-1}$. We also study marginal densities and the statistics of total losses.

\section{Simple mode} %
Let us consider a simple state with energy $\varepsilon_0$, which is coupled to the continuum by means of the decay amplitudes $A_c$, where index $c$ labels the scattering channels open at energy $E$. In the resonance approximation, $A_c$ may be assumed to be energy-independent, leading to a multichannel Breit-Wigner formula \cite{Mahaux}
$
  S^{(0)}_{ab}(E) = \delta_{ab} - iA_a^*A_b/(E-\varepsilon_0+\frac{i}{2}\Gamma_0)
$
for the $S$-matrix elements. The escape width $\Gamma_0$ is then given by the sum of the partial (per channel) widths, $\Gamma_0=\sum_{c}|A_c|^2$. This ensures the unitarity of the $S$ matrix (at real $E$).

Following \cite{Bohr,soko97}, an interaction between such a mode and the surrounding background described by a Hamiltonian $\Hbg$ results in the modified energy dependence of the $S$ matrix,
\begin{equation}\label{S_ab}
S_{ab}(E) = \delta_{ab} - i \frac{A_a^*A_b}{E-\varepsilon_0+\frac{i}{2}\Gamma_0 - g(E)}\,,
\end{equation}
where $g(E)=V^\dagger(E-\Hbg)^{-1}V$ is the strength function and $V$ stands for a coupling vector to $N$ background states. The latter usually have a very complex structure, fluctuating strongly on the scale of the mean level spacing $\Delta\sim1/N$.  When averaged over this fine structure, the mean amplitudes acquire an effective additional damping and read
$$
 \langle S_{ab}(E) \rangle = \delta_{ab} - i \frac{A_a^*A_b}{E-\varepsilon_0+\frac{i}{2}(\Gamma_0+\Gamma_{\downarrow})}\,,
$$
where
$\Gamma_{\downarrow}\equiv2\mathrm{Im}\langle{g(E-i0)}\rangle = 2\pi\|V\|^2/N\Delta$ is the spreading width. Introducing $\eta=\Gamma_{\downarrow}/\Gamma_0$ as a natural parameter controlling the strength of coupling to the background, we can cast the matrix $S$ at the resonance energy $E=\varepsilon_0$ as follows:
\begin{equation}\label{S}
  S = 1 - \frac{1}{1+i\eta K}(1-S^{(0)})\,.
\end{equation}
Here, $K\equiv2g(\varepsilon_0)/\Gamma_{\downarrow}$ has the meaning of the (dimensionless) local Green's function of the complex background \cite{fyod04b}. The unitary matrix $S^{(0)}$ stands for the deterministic part of $S$, $S^{(0)}_{ab} = \delta_{ab} - \frac{2}{\Gamma_0} A^*_aA_b$, accounting for the direct mixing of the channels. Expression (\ref{S}) provides the multichannel generalization of two-channel formulas derived recently in \cite{savi17}.

The established connection of $S$ to the background spectrum enables us to accommodate its physically relevant properties. Following the RMT paradigm \cite{guhr98,Stoeckmann}, we model $\Hbg$ by a random $N \times N$ matrix drawn from the Gaussian orthogonal (GOE) or unitary (GUE) ensemble, depending on the presence or absence of time-reversal invariance (TRI), respectively. Universal fluctuations are then expected to occur in the limit $N\gg1$. Furthermore,
homogeneous dissipation  can be easily taken into account by uniform broadening $\Gamma_\mathrm{abs}$ of the background states. Since such a damping is operationally equivalent \cite{savi03a} to the purely imaginary shift $\varepsilon_0+\frac{i}{2}\Gamma_\mathrm{abs}$ in the Green's function $K$, the latter becomes complex,
\begin{equation}\label{K}
  K = (2/\Gamma_{\downarrow}) g(\varepsilon_0+i\Gamma_\mathrm{abs}/2) \equiv u - iv,
\end{equation}
with $v>0$ being the local density of states (normalized as $\aver{v}=1$) \cite{fyod04b}. The universal statistics of mutually correlated random variables $u$ and $v$ is solely determined by the (dimensionless) absorption rate $\gamma = 2\pi \Gamma_\mathrm{abs}/\Delta$. They have the following joint probability density function (jpdf) \cite{fyod04b}:
\begin{align} \label{P(u,v)}
  \mathcal{P}(u,v) = \frac{1}{2\pi v^2}P_0(x),\quad x=\frac{u^2+v^2+1}{2v}>1\,.
\end{align}
In the present context, the function $P_0(x)$ has the meaning of the distribution of reflection induced by the background~\cite{note1}. This function is known exactly for both symmetry classes as well as in the crossover regime of gradually broken TRI \cite{savi05,fyod05}. We now apply these findings to derive nonperturbative results for the joint statistics of reflection and transmission of the simple mode at arbitrary values of $\eta$ and $\gamma$.

Scattering in a given channel `$a$' is commonly studied by means of the coefficients of reflection $R \equiv |S_{aa}|^2$ and total transmission $T \equiv \sum_{b\neq a}|S_{ab}|^2$. Making use of Eq.~(\ref{S}), one finds that these two quantities are expressed as follows
\begin{subequations}\label{R,T}
 \begin{eqnarray}
  R &=& \frac{(\Szero+\eta v)^2+\eta^2u^2}{(1 + \eta v)^2+\eta^2u^2}\,,
  \\
  T &=& \frac{1}{(1+\eta v)^2+\eta^2u^2}T_0\,,
 \end{eqnarray}
\end{subequations}
where $T_0\equiv\sum_{b\neq a}|S_{ab}^{(0)}|^2=1-(\Szero)^2$ is the total transmission coefficient in a ``clean'' system. At zero absorption, we have $v\equiv0$ and thus $R+T=1$ in agreement with the flux conservation. The latter is no longer valid at finite absorption, when $S$ becomes subunitary. Such a unitarity deficit can be naturally described by the following positive quantity:
\begin{equation}\label{delta}
  D \equiv 1-R-T = \frac{2(1-S^{(0)}_{aa})\eta v}{(1 + \eta v)^2+\eta^2u^2}
  \leq \frac{1-\Szero}{2},
\end{equation}
which gives the part of the total flux in the channel that gets dissipated in the background. The deficit $D=0$ identically at $\Szero=1$, when the channel is closed. It covers its maximum range $0\leq{D}\leq1$ at $\Szero=-1$, when the wave gets reflected  in full after the interaction with the background. (Note that both cases correspond to zero transmission.) We will study the probability distribution of $D$ below as well.

\section{Joint distribution of $R$ and $T$} %
Relations (\ref{R,T}) enable us to relate the jpdf in question to that from (\ref{P(u,v)}), expressing the result in terms of the known function $P_0(x)$. Formally, this amounts to computing the relevant Jacobians. Conceptually, such a link provides a duality between two different viewpoints when studying statistical fluctuations either ``from outside'' (e.g., via scattering measures \cite{kuhl13,kuhl05,koeb10,diet10}) or ``from inside'' (e.g., impedance and admittance \cite{hemm05,hemm06, grad14}).

\subsection{Perfect coupling}
\begin{figure}
  \centering
  \includegraphics[width=.95\linewidth]{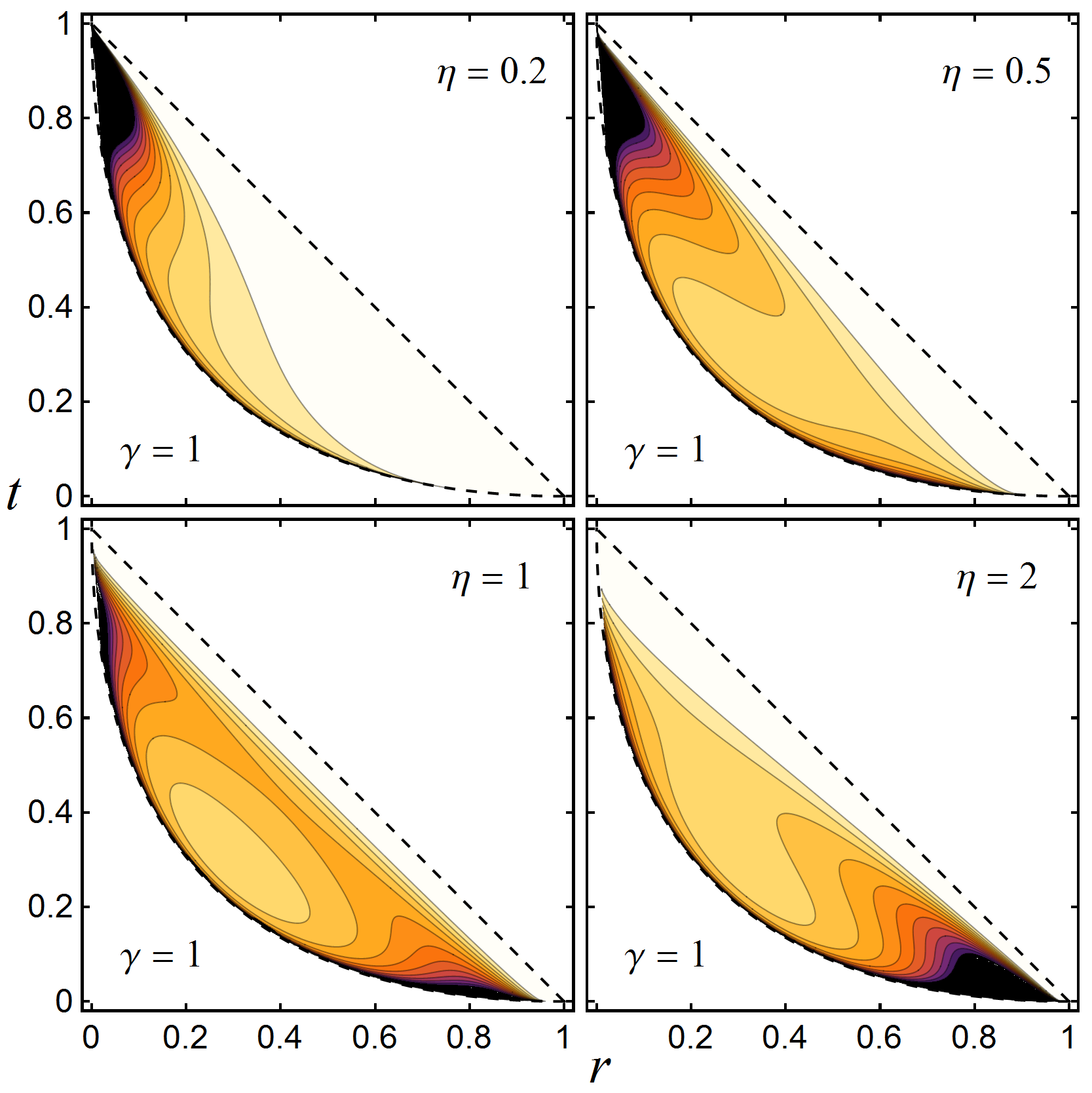}
  \caption{Contour plots of the joint distribution (\ref{P(r,t)}) of reflection and transmission for the GUE case of the chaotic background at moderate absorption $\gamma=1$ and various couplings $\eta$. Darker regions correspond to higher values of the jpdf, whereas the dashed lines indicate the boundaries of the distribution support. Note the symmetry of the distribution at $\eta=1$ and at the reciprocal values of $\eta$.
  \label{fig1} }
\end{figure}

It is instructive to consider first the case of perfect coupling, $\Szero=0$ ($T_0=1$). We reserve the notation $t=T|_{T_0=1}$ and $r=R|_{T_0=1}$. Computing the Jacobian $\frac{\partial(u,v)}{\partial(r,t)}$, we find after some algebra the following attractive formula for the jpdf:
\begin{equation}\label{P(r,t)}
  \mathcal{P}(r,t) = \frac{2}{\pi(1-r-t)^2\sqrt{y}}
           P_0\biggl(\frac{\eta^{-1}r+\eta t}{1-r-t}\biggr)\,,
\end{equation}
for $1-r-t>0$ and $y\equiv1+2rt-(1-r)^2-(1-t)^2>0$, being zero otherwise. It follows at once that function (\ref{P(r,t)}) has the following important symmetry under $\eta\to\eta^{-1}$:
\begin{equation}\label{sym_joint}
  \mathcal{P}(r,t)|_{\eta} = \mathcal{P}(t,r)|_{\eta^{-1}}\,.
\end{equation}
This holds at any $\gamma$ and shows that the background coupling $\eta$ controls the weight of the total flux distribution between its reflection and transmission sectors. In particular, distribution (\ref{P(r,t)}) becomes symmetric with respect to the line $r=t$ at the special coupling $\eta=1$. This is further illustrated on Fig.~\ref{fig1}.

In the limit of vanishing absorption, $\gamma\to0$, the function  $P_0(x)$ is known \cite{fyod04b} to reduce to $\delta(\frac{1}{x})$, which readily yields
\begin{equation}\label{P(r,t)gam0}
  \mathcal{P}_{\gamma=0}(r,t) =  \delta(1-r-t) \mathcal{P}_{0}(t).
\end{equation}
The first (singular) factor above stands for the conditional probability density function (pdf) of $r$, expressing here the flux conservation. The marginal distribution $\mathcal{P}_0(t)$ denotes the transmission distribution for an ideal stable background (at zero absorption) and reads \cite{savi17}
\begin{align}\label{PT_stable}
  \mathcal{P}_0(t)= \frac{1}{\pi\sqrt{t(1-t)} [\eta t+\eta^{-1}(1-t)]}\,.
\end{align}
This distribution is insensitive to the presence of TRI.

The singularity of the joint distribution is removed at finite absorption, since $r$ and $t$ are no longer functionally dependent. Further analysis is possible in the physically interesting limiting cases of weak and strong  absorption, when the function $P_0$ is known to take simpler asymptotic forms \cite{fyod04b}. One has $P_0(x)\approx\frac{(\beta\gamma)^{\beta/2+1}}{4\Gamma(\beta/2+1)} (\frac{x+1}{4})^{\beta/2}\,e^{-\frac{\beta\gamma}{4}(x+1)}$ at $\gamma\ll1$, where $\beta=1$ ($\beta=2$) stands for the GOE (GUE) case. As a result, the jpdf is concentrated within a thin layer near the boundary $r=1-t$, being approximated in the leading order by
\begin{align}\label{jpdf_weak}
  \mathcal{P}_{\gamma\ll1}(r,t)\approx \frac{1}{\Gamma(\frac{\beta}{2}+1)d}
  \left(\frac{\beta\gamma z}{4d}\right)^{\frac{\beta}{2}+1}
  e^{-\frac{\beta\gamma z}{4d}}\mathcal{P}_0(t) \,,
\end{align}
where $z\equiv\eta t+\eta^{-1}(1-t)$ and $d\equiv1-r-t$ (note $d\sim\gamma\ll1$). When $\gamma$ is increased, the distribution starts exploring its whole support. Making use of $P_0(x)\approx\frac{\beta\gamma}{4} e^{-\frac{\beta\gamma}{4}(x-1)}$ at $\gamma\gg1$, we find an asymptotic expression at strong absorption,
\begin{align}\label{jpdf_strong}
  \mathcal{P}_{\gamma\gg1}(r,t)\approx
  \frac{\beta\gamma \exp\left[-\frac{\beta\gamma}{4}\frac{(1+\eta^{-1})r+(1+\eta)t-1}{1-r-t}\right]
  }{ 2\pi(1-r-t)^2\sqrt{y}}\,.
\end{align}
This clearly shows that the transmission and reflection exhibit nontrivial statistical correlations even at large absorption.

\subsection{Marginal distributions}
The marginal distributions of transmission or reflection can now be obtained from the jpdf (\ref{P(r,t)}) by integrating it over $r$ or $t$. Note that $y$ is symmetric under the interchange $r\leftrightarrow t$ and thus can be also cast as follows:
\begin{equation}\label{y}
  y = (r_{+}-r)(r-r_{-}) = (t_{+}-t)(t-t_{-}),
\end{equation}
with $r_{\pm}=(1\pm\sqrt{t})^2$ and $t_{\pm}=(1\pm\sqrt{r})^2$. One readily finds the following expression for the transmission distribution:
\begin{equation}\label{Ptr}
  \mathcal{P}^{\mathrm{(tr)}}_{\eta}(t) = \int_{r_-}^{1-t}\!\!\frac{dr}{\pi(1-r-t)^2}
  \frac{2P_0\bigl(\frac{\eta^{-1}r+\eta t}{1-r-t}\bigr) }{ \sqrt{(r_{+}-r)(r-r_{-})} }\,,
\end{equation}
for $0\le{t}\le1$ and zero otherwise \cite{note2}.   The advantage of representation (\ref{Ptr}) is that it utilizes the symmetry property (\ref{sym_joint}) explicitly. It becomes then obvious that the distribution of reflection is simply related to that of transmission as follows
\begin{equation}\label{Pref}
  \mathcal{P}^{\mathrm{(ref)}}_{\eta}(r) = \mathcal{P}^{\mathrm{(tr)}}_{\eta^{-1}}(r)\,.
\end{equation}
This is a remarkable relation showing that despite lacking any apparent connection between the reflection and transmission coefficients at finite absorption, their distribution functions turn out to be linked by symmetry (\ref{Pref}). With explicit formulas for $P_0$ found in \cite{savi05,fyod05}, Eqs.~(\ref{P(r,t)}), (\ref{Ptr}) and (\ref{Pref}) provide the exact solution to the problem at arbitrary $\eta$ and $\gamma$.

\begin{figure}
  \centering
  \includegraphics[width=\linewidth]{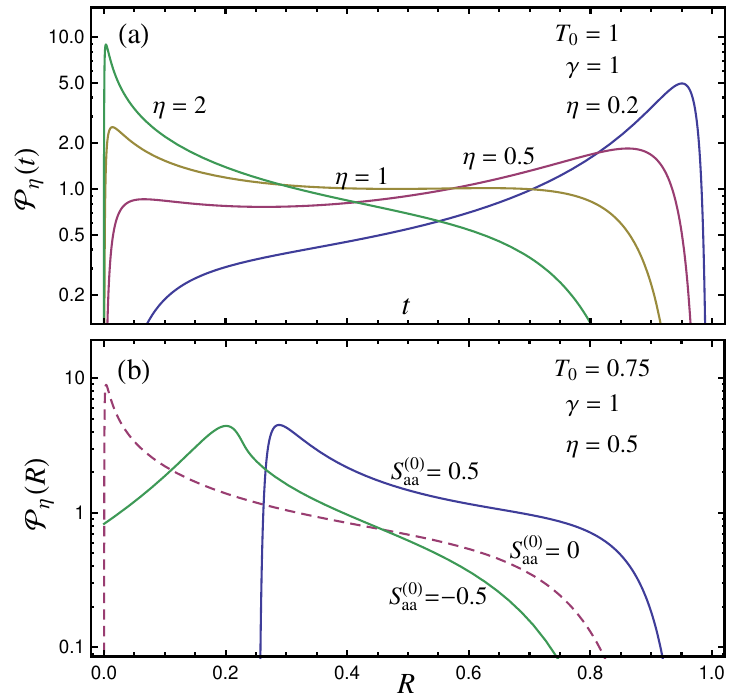}
  \caption{Distribution of transmission (top) and reflection (bottom) for the GUE case at $\gamma=1$ and various $\eta$.
  (a) Transmission distribution (\ref{Ptr}) at perfect coupling $T_0=1$. By the symmetry (\ref{Pref}), the corresponding reflection distributions would be given by the same curves at the reciprocal values of $\eta$.
  (b) Reflection distribution (\ref{P(R)}) at nonperfect coupling $T_0=0.75$, corresponding to $\Szero=\pm\sqrt{1-T_0}$. Note a hard gap $R>1-T_0$ of the distribution when $\Szero>0$ and a finite value of $\mathcal{P}(0)$ when $\Szero<0$. The dashed line shows the reflection distribution at perfect coupling for comparison.
  \label{fig2} }
\end{figure}

In the limiting cases of weak and strong  absorption, one can further make use of the asymptotic forms derived above. Performing the integration in (\ref{Ptr}), we arrive after some algebra at the following leading-order result at $\gamma\ll1$:
\begin{equation}\label{Ptr_weak}
  \mathcal{P}^{\mathrm{(tr)}}_{\gamma\ll1}(t) \approx \mathcal{P}_{0}(t)
  \exp\left[-\frac{\beta\gamma}{8\eta}\frac{(1+(\eta-1)\sqrt{t})^2}{\sqrt{t}(1-\sqrt{t})}\right]\,.
\end{equation}
It has a bimodal profile of (\ref{PT_stable}) in the bulk, which gets crucially modified near the edges due to exponential cutoffs induced by absorption. In the opposite case of $\gamma\gg1$, one finds
\begin{equation}\label{Pgam_strong}
    \mathcal{P}^{\mathrm{(tr)}}_{\gamma\gg1}(t)\approx
     \frac{ \sqrt{\beta\gamma}
      \exp\left[-\frac{\beta\gamma}{8\eta}\frac{(1-(\eta+1)\sqrt{t})^2}{\sqrt{t}(1-\sqrt{t})}\right]
     }{ 4\sqrt{\pi}t^{3/4}(1-\sqrt{t})\sqrt{\eta t + \eta^{-1}(1-t)} }.
\end{equation}
Figure~\ref{fig2} shows $\mathcal{P}^{\mathrm{(tr)}}_{\eta}(t)$ at moderate absorption $\gamma=1$.

\subsection{Nonperfect coupling} %
In the general case of $\Szero\neq0$, it is also convenient to express the reflection and transmission coefficients (\ref{R,T}) in terms of $r$ and $t$ studied above. One finds
\begin{equation}\label{RT-rt}
  T=T_0t, \quad R = \Szero + (1-\Szero)(r-\Szero t).
\end{equation}
Now only a part (given by $T_0$) of the incoming flux contributes to the transmission. Thus the distribution of $T$ is obtained by a simple rescaling of expression (\ref{Ptr}). The reflection coefficient takes a more elaborate form because of the interference between the two reflected waves, the one backscattered directly at the channel interface and the one originating from the background. The corresponding distribution can be found in a closed form using Eqs.~(\ref{P(r,t)}) and (\ref{RT-rt}) and reads
\begin{equation}\label{P(R)}
  \mathcal{P}_{\eta}(R) = \int_{T_-}^{T_*}\!\! \frac{dT}{\pi(1{-}R{-}T)^2}\frac{2P_0(X)}{\sqrt{(T_{+}-T)(T-T_{-})}}\,,
\end{equation}
where $T_*=\min(1-R,T_+)$,  $T_{\pm}=\frac{1+\Szero}{1-\Szero}(1\pm\sqrt{R})^2$, and
\begin{equation}\label{x}
  X = \frac{(1+\Szero)(R-\Szero)+T(\eta^2+\Szero)}{\eta(1+\Szero)(1-R-T)}\,.
\end{equation}
It reduces to Eq.~(\ref{Pref}) at perfect coupling, $\Szero=0$.

A particular feature of the reflection distribution (\ref{P(R)}) is the dependence of its support on the sign of $\Szero$ (see Fig.~\ref{fig2}(b) and Ref.~\cite{savi17}). The distribution vanishes identically for $R\leq1-T_0$, when $\Szero>0$, and covers the whole range $0\leq R\leq1$, when $\Szero<0$. This follows from the compatibility condition $T_-<T_*$ and is, of course, in agreement with definition (\ref{R,T}).

\section{Application to loss statistics} %
We now apply the obtained results to discuss the distribution of the unitary deficit (\ref{delta}), which is a useful measure of total losses in the system. By the construction $D=(1-\Szero)d$, where $d=D|_{T_0=1}=1-r-t$ is the deficit at perfect coupling \cite{note3}. We note that property~(\ref{sym_joint}) enforces the deficit distribution to depend on $\eta$ and $\eta^{-1}$ in a symmetric way. It is therefore convenient to introduce $g\equiv\frac{1}{2}(\eta+\frac{1}{\eta})\ge1$ as the effective coupling constant to the background. After some algebra, we arrive at the following exact result for the distribution of the loss parameter $d$ ($0\leq2d\leq1$):
\begin{equation}\label{Ploss}
  \mathcal{P}_{g}(d) = \int_{0}^{\pi}\frac{d\theta}{\pi d^2}
  P_0\biggl[\frac{g(1-d)}{d} - \frac{\sqrt{(g^2-1)(1-2d)}}{d}\cos\theta\biggr].
\end{equation}
At the special coupling $g=1$ ($\eta=1$), this expression simplifies further to $\mathcal{P}_{g=1}(d)=\frac{1}{d^2}P_0(\frac{1-d}{d})$. The asymptotic forms of $\mathcal{P}_{g}(d)$ can also be obtained in the limits of weak and strong absorption. In particular, in the latter case it reads
\begin{equation}\label{PlossStr}
  \mathcal{P}_{\gamma\gg1}(d) \approx
  \frac{\beta\gamma e^{\frac{\beta\gamma}{4}(g+1-\frac{g}{d})} }{ 4d^2 }
  I_0 \biggl[ \frac{\beta\gamma}{4d}\sqrt{(g^2-1)(1-2d)} \biggr],
\end{equation}
where $I_0(x)$ is a modified Bessel function.

\begin{figure}
  \centering
  \includegraphics[width=\linewidth]{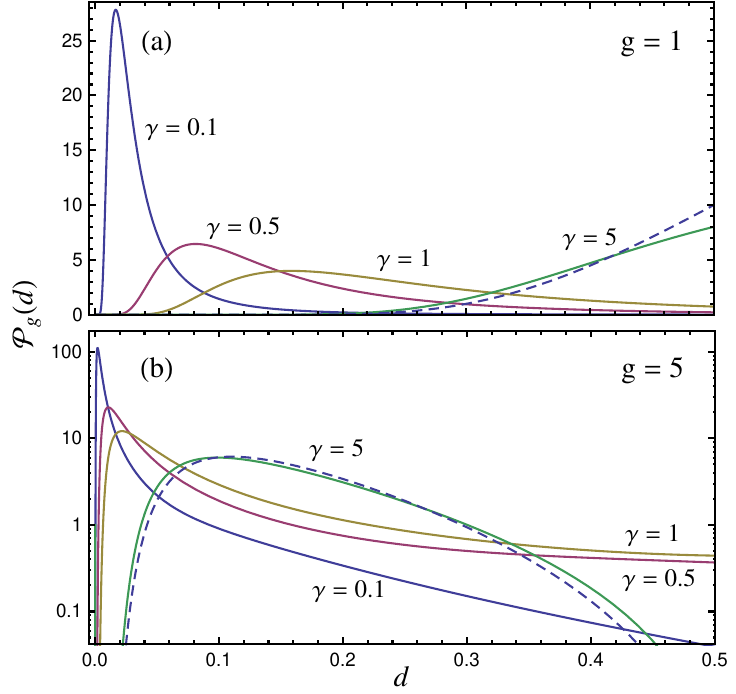}
  \caption{Distribution (\ref{Ploss}) of the loss parameter $d=1-r-t$ (unitarity deficit) for the GUE case at the increased absorption $\gamma$. The background coupling $g=1$ (top) and $g=5$ (bottom, note the semilog scale in this case). The dashed lines at $\gamma=5$ correspond to the asymptotic expression (\ref{PlossStr}) at strong absorption.
  \label{fig3} }
\end{figure}

For arbitrary absorption, expression (\ref{Ploss}) can be evaluated further only in the GUE case, when $P_0$ takes the following simple form: $P_0(x) = \frac{1}{2}[\frac{\gamma}{2}(x+1)A + B]e^{-\gamma(x+1)/2}$, with the $\gamma$-dependent constants $A= e^{\gamma}-1$ and $B=1+\gamma-e^{\gamma}$~\cite{fyod04b,been01}. Performing the subsequent integration results in
\begin{equation}\label{PlossGUE}
  \mathcal{P}^{\mathrm{(gue)}}_{g}(d) = \frac{1}{2d^2} \left(B-A\gamma\frac{\partial}{\partial\gamma}\right) F\,,
\end{equation}
where $F= e^{\frac{\gamma}{2}(g-1-\frac{g}{d})} I_0 \bigl[ \frac{\gamma}{2d}\sqrt{(g^2-1)(1-2d)} \bigr]$. This distribution for various values of $\gamma$ and $g$ is shown on Fig.~\ref{fig3}. The behavior of $\mathcal{P}_{g}(d)$ in the GOE case is similar and can be qualitatively described by rescaling $\gamma\to\frac{\gamma}{2}$ in (\ref{PlossGUE}).

It is worth noting that the unitarity deficit is closely related to the time-delay matrix $Q$ at finite absorption \cite{savi03a} as well as to the so-called probability of no return $\tau\equiv1-R$ \cite{fyod03}. The former is defined by $Q=\Gamma^{-1}_\mathrm{abs}(1-S^{\dag}S)$, yielding  $D=\Gamma_\mathrm{abs}Q_{aa}$, whereas the later is given by $\tau=D+T$. Refs.~\cite{savi03a,fyod03} provide the exact multichannel formulae for the mean eigenvalue density of $Q$ and for the distribution of $\tau$ in a different setting of fully chaotic scattering without a direct mixing between the channels. The two distributions are distinct in general, but reduce to the same expression in the special case of the single channel, since $T=0$ then identically. It turns out that the deficit distribution (\ref{Ploss}) is also given by the very same expression, provided that $d=\tau/2$ and $\eta$ is identified with the channel coupling. This can be substantiated by noting that zero transmission in the present model is realized at $\Szero=-1$, resulting in $D=2d$ and $S_{aa} = \frac{1-i\eta K}{1+i\eta K}$. The latter is the usual form for the elastic (single-channel) scattering \cite{verb85}, with $\eta$ now playing the role of coupling to the continuum. This proves the connection observed.

\section{Discussion and Conclusion} %
The approach developed shows that scattering on the simple mode coupled to the complex background serves as a sensitive probe of its internal structure. Fluctuations in scattering induced by the background states are governed by the interplay between the spreading width and losses in the environment. We have derived the joint distribution of the reflection and total transmission at arbitrary coupling to and absorption in the background. The reflection and transmission coefficients are found to develop strong and nontrivial statistical correlations, which remain essential even in the limit of strong absorption. The obtained results have been further applied to derive and study the exact statistics of total losses in the system.

Of particularly interest is the new and remarkable symmetry between fluctuations in reflection and transmission sectors, expressed by Eqs.~(\ref{sym_joint}) and (\ref{Pref}), which holds at arbitrary absorption. This can be traced back to the symmetry properties of the local density of states, which were first established for ergodic states in \cite{fyod04b,savi05} and then generalized to multifractal spectra at Anderson transition \cite{mirl06} and at critical points of other disordered systems \cite{gruz11}. Studying the symmetry of relevant multifractal exponents has recently become accessible experimentally \cite{khay15}. The formalism presented here offers the promising way to study manifestations of such symmetries at the level of scattering characteristics.

The approach is flexible in incorporating other real-world effects, e.g., inhomogeneous losses following \cite{savi06b}. The two relevant model parameters, the background coupling ($\eta$) and absorption strength ($\gamma$), can be extracted from scattering data as discussed in Ref.~\cite{savi17appa}. We also note that our results are applicable equally well for the systems with as well as without time-reversal invariance. The latter (and experimentally more challenging) case has been recently realised using microwave graphs \cite{lawn19}. Thus one can expect further applications of our findings within a broader context of wave chaotic systems.
\vspace*{2ex}
\acknowledgments
\vspace*{-2ex}
I am grateful to Y. Fyodorov, U. Kuhl, O. Legrand, F. Mortessagne, and M. Richter for useful discussions of various aspects of chaotic scattering relevant to this work.

%\bibliography{../Bib/refs,../Bib/books,simple_add}

\begin{thebibliography}{43}%
\makeatletter
\providecommand \@ifxundefined [1]{%
 \@ifx{#1\undefined}
}%
\providecommand \@ifnum [1]{%
 \ifnum #1\expandafter \@firstoftwo
 \else \expandafter \@secondoftwo
 \fi
}%
\providecommand \@ifx [1]{%
 \ifx #1\expandafter \@firstoftwo
 \else \expandafter \@secondoftwo
 \fi
}%
\providecommand \natexlab [1]{#1}%
\providecommand \enquote  [1]{``#1''}%
\providecommand \bibnamefont  [1]{#1}%
\providecommand \bibfnamefont [1]{#1}%
\providecommand \citenamefont [1]{#1}%
\providecommand \href@noop [0]{\@secondoftwo}%
\providecommand \href [0]{\begingroup \@sanitize@url \@href}%
\providecommand \@href[1]{\@@startlink{#1}\@@href}%
\providecommand \@@href[1]{\endgroup#1\@@endlink}%
\providecommand \@sanitize@url [0]{\catcode `\\12\catcode `\$12\catcode
  `\&12\catcode `\#12\catcode `\^12\catcode `\_12\catcode `\%12\relax}%
\providecommand \@@startlink[1]{}%
\providecommand \@@endlink[0]{}%
\providecommand \url  [0]{\begingroup\@sanitize@url \@url }%
\providecommand \@url [1]{\endgroup\@href {#1}{\urlprefix }}%
\providecommand \urlprefix  [0]{URL }%
\providecommand \Eprint [0]{\href }%
\providecommand \doibase [0]{https://doi.org/}%
\providecommand \selectlanguage [0]{\@gobble}%
\providecommand \bibinfo  [0]{\@secondoftwo}%
\providecommand \bibfield  [0]{\@secondoftwo}%
\providecommand \translation [1]{[#1]}%
\providecommand \BibitemOpen [0]{}%
\providecommand \bibitemStop [0]{}%
\providecommand \bibitemNoStop [0]{.\EOS\space}%
\providecommand \EOS [0]{\spacefactor3000\relax}%
\providecommand \BibitemShut  [1]{\csname bibitem#1\endcsname}%
\let\auto@bib@innerbib\@empty
%</preamble>
\bibitem [{\citenamefont {Sokolov}\ and\ \citenamefont
  {Zelevinsky}(1997)}]{soko97}%
  \BibitemOpen
  \bibfield  {author} {\bibinfo {author} {\bibfnamefont {V.~V.}\ \bibnamefont
  {Sokolov}}\ and\ \bibinfo {author} {\bibfnamefont {V.}~\bibnamefont
  {Zelevinsky}},\ }\bibfield  {title} {\bibinfo {title} {Simple mode on a
  highly excited background: Collective strength and damping in the
  continuum},\ }\href {https://doi.org/10.1103/PhysRevC.56.311} {\bibfield
  {journal} {\bibinfo  {journal} {Phys. Rev. C}\ }\textbf {\bibinfo {volume}
  {56}},\ \bibinfo {pages} {311} (\bibinfo {year} {1997})}\BibitemShut
  {NoStop}%
\bibitem [{\citenamefont {Bohr}\ and\ \citenamefont {Mottelson}(1969)}]{Bohr}%
  \BibitemOpen
  \bibfield  {author} {\bibinfo {author} {\bibfnamefont {A.}~\bibnamefont
  {Bohr}}\ and\ \bibinfo {author} {\bibfnamefont {B.~R.}\ \bibnamefont
  {Mottelson}},\ }\href@noop {} {\emph {\bibinfo {title} {Nuclear Structure}}}\
  (\bibinfo  {publisher} {Benjamin},\ \bibinfo {address} {New York},\ \bibinfo
  {year} {1969})\BibitemShut {NoStop}%
\bibitem [{\citenamefont {Harney}\ \emph {et~al.}(1986)\citenamefont {Harney},
  \citenamefont {Richter},\ and\ \citenamefont {Weidenm\"uller}}]{harn86}%
  \BibitemOpen
  \bibfield  {author} {\bibinfo {author} {\bibfnamefont {H.~L.}\ \bibnamefont
  {Harney}}, \bibinfo {author} {\bibfnamefont {A.}~\bibnamefont {Richter}},\
  and\ \bibinfo {author} {\bibfnamefont {H.~A.}\ \bibnamefont
  {Weidenm\"uller}},\ }\bibfield  {title} {\bibinfo {title} {Breaking of
  isospin symmetry in compound-nucleus reactions},\ }\href
  {https://doi.org/10.1103/RevModPhys.58.607} {\bibfield  {journal} {\bibinfo
  {journal} {Rev. Mod. Phys.}\ }\textbf {\bibinfo {volume} {58}},\ \bibinfo
  {pages} {607} (\bibinfo {year} {1986})}\BibitemShut {NoStop}%
\bibitem [{\citenamefont {Sokolov}\ \emph {et~al.}(1997)\citenamefont
  {Sokolov}, \citenamefont {Rotter}, \citenamefont {Savin},\ and\ \citenamefont
  {M{\"{u}}ller}}]{soko97i}%
  \BibitemOpen
  \bibfield  {author} {\bibinfo {author} {\bibfnamefont {V.~V.}\ \bibnamefont
  {Sokolov}}, \bibinfo {author} {\bibfnamefont {I.}~\bibnamefont {Rotter}},
  \bibinfo {author} {\bibfnamefont {D.~V.}\ \bibnamefont {Savin}},\ and\
  \bibinfo {author} {\bibfnamefont {M.}~\bibnamefont {M{\"{u}}ller}},\
  }\bibfield  {title} {\bibinfo {title} {Interfering doorway states and giant
  resonances. {I}. {R}esonance spectrum and multipole strengths},\ }\href
  {https://doi.org/10.1103/PhysRevC.56.1031} {\bibfield  {journal} {\bibinfo
  {journal} {Phys. Rev. C}\ }\textbf {\bibinfo {volume} {56}},\ \bibinfo
  {pages} {1031} (\bibinfo {year} {1997})}\BibitemShut {NoStop}%
\bibitem [{\citenamefont {Gu}\ and\ \citenamefont
  {Weidenm\"uller}(1999)}]{gu99}%
  \BibitemOpen
  \bibfield  {author} {\bibinfo {author} {\bibfnamefont {J.-Z.}\ \bibnamefont
  {Gu}}\ and\ \bibinfo {author} {\bibfnamefont {H.}~\bibnamefont
  {Weidenm\"uller}},\ }\bibfield  {title} {\bibinfo {title} {Decay out of a
  superdeformed band},\ }\href {https://doi.org/10.1016/S0375-9474(99)00362-0}
  {\bibfield  {journal} {\bibinfo  {journal} {Nucl. Phys. A}\ }\textbf
  {\bibinfo {volume} {660}},\ \bibinfo {pages} {197} (\bibinfo {year}
  {1999})}\BibitemShut {NoStop}%
\bibitem [{\citenamefont {Zelevinsky}\ and\ \citenamefont
  {Volya}(2016)}]{zele16}%
  \BibitemOpen
  \bibfield  {author} {\bibinfo {author} {\bibfnamefont {V.}~\bibnamefont
  {Zelevinsky}}\ and\ \bibinfo {author} {\bibfnamefont {A.}~\bibnamefont
  {Volya}},\ }\bibfield  {title} {\bibinfo {title} {Chaotic features of nuclear
  structure and dynamics: selected topics},\ }\href
  {https://doi.org/10.1088/0031-8949/91/3/033006} {\bibfield  {journal}
  {\bibinfo  {journal} {Phys. Scr.}\ }\textbf {\bibinfo {volume} {91}},\
  \bibinfo {pages} {033006} (\bibinfo {year} {2016})}\BibitemShut {NoStop}%
\bibitem [{\citenamefont {Aberg}\ \emph {et~al.}(2008)\citenamefont {Aberg},
  \citenamefont {Guhr}, \citenamefont {{Miski-Oglu}},\ and\ \citenamefont
  {Richter}}]{aber08}%
  \BibitemOpen
  \bibfield  {author} {\bibinfo {author} {\bibfnamefont {S.}~\bibnamefont
  {Aberg}}, \bibinfo {author} {\bibfnamefont {T.}~\bibnamefont {Guhr}},
  \bibinfo {author} {\bibfnamefont {M.}~\bibnamefont {{Miski-Oglu}}},\ and\
  \bibinfo {author} {\bibfnamefont {A.}~\bibnamefont {Richter}},\ }\bibfield
  {title} {\bibinfo {title} {Superscars in billiards: {A} model for doorway
  states in quantum spectra},\ }\href
  {https://doi.org/10.1103/PhysRevLett.100.204101} {\bibfield  {journal}
  {\bibinfo  {journal} {Phys. Rev. Lett.}\ }\textbf {\bibinfo {volume} {100}},\
  \bibinfo {pages} {204101} (\bibinfo {year} {2008})}\BibitemShut {NoStop}%
\bibitem [{\citenamefont {Sokolov}(2010)}]{soko10}%
  \BibitemOpen
  \bibfield  {author} {\bibinfo {author} {\bibfnamefont {V.~V.}\ \bibnamefont
  {Sokolov}},\ }\bibfield  {title} {\bibinfo {title} {Ballistic electron
  quantum transport in the presence of a disordered background},\ }\href
  {https://doi.org/10.1088/1751-8113/43/26/265102} {\bibfield  {journal}
  {\bibinfo  {journal} {J. Phys. A}\ }\textbf {\bibinfo {volume} {43}},\
  \bibinfo {pages} {265102} (\bibinfo {year} {2010})}\BibitemShut {NoStop}%
\bibitem [{\citenamefont {Morales}\ \emph {et~al.}(2012)\citenamefont
  {Morales}, \citenamefont {de~Anda}, \citenamefont {Flores}, \citenamefont
  {Guti\'errez}, \citenamefont {M\'endez-S\'anchez}, \citenamefont
  {Monsivais},\ and\ \citenamefont {Mora}}]{mora12}%
  \BibitemOpen
  \bibfield  {author} {\bibinfo {author} {\bibfnamefont {A.}~\bibnamefont
  {Morales}}, \bibinfo {author} {\bibfnamefont {A.~D.}\ \bibnamefont
  {de~Anda}}, \bibinfo {author} {\bibfnamefont {J.}~\bibnamefont {Flores}},
  \bibinfo {author} {\bibfnamefont {L.}~\bibnamefont {Guti\'errez}}, \bibinfo
  {author} {\bibfnamefont {R.~A.}\ \bibnamefont {M\'endez-S\'anchez}}, \bibinfo
  {author} {\bibfnamefont {G.}~\bibnamefont {Monsivais}},\ and\ \bibinfo
  {author} {\bibfnamefont {P.}~\bibnamefont {Mora}},\ }\bibfield  {title}
  {\bibinfo {title} {Doorway states in quasi-one-dimensional elastic systems},\
  }\href {https://doi.org/10.1209/0295-5075/99/54002} {\bibfield  {journal}
  {\bibinfo  {journal} {Europhys. Lett.}\ }\textbf {\bibinfo {volume} {99}},\
  \bibinfo {pages} {54002} (\bibinfo {year} {2012})}\BibitemShut {NoStop}%
\bibitem [{\citenamefont {Savin}\ \emph {et~al.}(2017)\citenamefont {Savin},
  \citenamefont {Richter}, \citenamefont {Kuhl}, \citenamefont {Legrand},\ and\
  \citenamefont {Mortessagne}}]{savi17}%
  \BibitemOpen
  \bibfield  {author} {\bibinfo {author} {\bibfnamefont {D.~V.}\ \bibnamefont
  {Savin}}, \bibinfo {author} {\bibfnamefont {M.}~\bibnamefont {Richter}},
  \bibinfo {author} {\bibfnamefont {U.}~\bibnamefont {Kuhl}}, \bibinfo {author}
  {\bibfnamefont {O.}~\bibnamefont {Legrand}},\ and\ \bibinfo {author}
  {\bibfnamefont {F.}~\bibnamefont {Mortessagne}},\ }\bibfield  {title}
  {\bibinfo {title} {Fluctuations in an established transmission in the
  presence of a complex environment},\ }\href
  {https://doi.org/10.1103/PhysRevE.96.032221} {\bibfield  {journal} {\bibinfo
  {journal} {Phys. Rev. E}\ }\textbf {\bibinfo {volume} {96}},\ \bibinfo
  {pages} {032221} (\bibinfo {year} {2017})}\BibitemShut {NoStop}%
\bibitem [{\citenamefont {Kuhl}\ \emph
  {et~al.}(2005{\natexlab{a}})\citenamefont {Kuhl}, \citenamefont
  {St{\"{o}}ckmann},\ and\ \citenamefont {Weaver}}]{kuhl05a}%
  \BibitemOpen
  \bibfield  {author} {\bibinfo {author} {\bibfnamefont {U.}~\bibnamefont
  {Kuhl}}, \bibinfo {author} {\bibfnamefont {H.-J.}\ \bibnamefont
  {St{\"{o}}ckmann}},\ and\ \bibinfo {author} {\bibfnamefont {R.}~\bibnamefont
  {Weaver}},\ }\bibfield  {title} {\bibinfo {title} {Classical wave experiments
  on chaotic scattering},\ }\href {https://doi.org/10.1088/0305-4470/38/49/001}
  {\bibfield  {journal} {\bibinfo  {journal} {J. Phys. A}\ }\textbf {\bibinfo
  {volume} {38}},\ \bibinfo {pages} {10433} (\bibinfo {year}
  {2005}{\natexlab{a}})}\BibitemShut {NoStop}%
\bibitem [{\citenamefont {Savin}\ and\ \citenamefont
  {Sommers}(2003)}]{savi03a}%
  \BibitemOpen
  \bibfield  {author} {\bibinfo {author} {\bibfnamefont {D.~V.}\ \bibnamefont
  {Savin}}\ and\ \bibinfo {author} {\bibfnamefont {H.-J.}\ \bibnamefont
  {Sommers}},\ }\bibfield  {title} {\bibinfo {title} {Delay times and
  reflection in chaotic cavities with absorption},\ }\href
  {https://doi.org/10.1103/PhysRevE.68.036211} {\bibfield  {journal} {\bibinfo
  {journal} {Phys. Rev. E}\ }\textbf {\bibinfo {volume} {68}},\ \bibinfo
  {pages} {036211} (\bibinfo {year} {2003})}\BibitemShut {NoStop}%
\bibitem [{\citenamefont {Fyodorov}(2003)}]{fyod03}%
  \BibitemOpen
  \bibfield  {author} {\bibinfo {author} {\bibfnamefont {Y.~V.}\ \bibnamefont
  {Fyodorov}},\ }\bibfield  {title} {\bibinfo {title} {Induced vs. spontaneous
  breakdown of {$S$}-matrix unitarity: Probability of no return in quantum
  chaotic and disordered systems},\ }\href {https://doi.org/10.1134/1.1622041}
  {\bibfield  {journal} {\bibinfo  {journal} {JETP Lett.}\ }\textbf {\bibinfo
  {volume} {78}},\ \bibinfo {pages} {250} (\bibinfo {year} {2003})}\BibitemShut
  {NoStop}%
\bibitem [{\citenamefont {Fyodorov}\ \emph {et~al.}(2005)\citenamefont
  {Fyodorov}, \citenamefont {Savin},\ and\ \citenamefont {Sommers}}]{fyod05}%
  \BibitemOpen
  \bibfield  {author} {\bibinfo {author} {\bibfnamefont {Y.~V.}\ \bibnamefont
  {Fyodorov}}, \bibinfo {author} {\bibfnamefont {D.~V.}\ \bibnamefont
  {Savin}},\ and\ \bibinfo {author} {\bibfnamefont {H.-J.}\ \bibnamefont
  {Sommers}},\ }\bibfield  {title} {\bibinfo {title} {Scattering, reflection
  and impedance of waves in chaotic and disordered systems with absorption},\
  }\href {https://doi.org/10.1088/0305-4470/38/49/017} {\bibfield  {journal}
  {\bibinfo  {journal} {J. Phys. A}\ }\textbf {\bibinfo {volume} {38}},\
  \bibinfo {pages} {10731} (\bibinfo {year} {2005})}\BibitemShut {NoStop}%
\bibitem [{\citenamefont {Kumar}\ \emph {et~al.}(2013)\citenamefont {Kumar},
  \citenamefont {Nock}, \citenamefont {Sommers}, \citenamefont {Guhr},
  \citenamefont {Dietz}, \citenamefont {Miski-Oglu}, \citenamefont {Richter},\
  and\ \citenamefont {Sch\"afer}}]{kuma13}%
  \BibitemOpen
  \bibfield  {author} {\bibinfo {author} {\bibfnamefont {S.}~\bibnamefont
  {Kumar}}, \bibinfo {author} {\bibfnamefont {A.}~\bibnamefont {Nock}},
  \bibinfo {author} {\bibfnamefont {H.-J.}\ \bibnamefont {Sommers}}, \bibinfo
  {author} {\bibfnamefont {T.}~\bibnamefont {Guhr}}, \bibinfo {author}
  {\bibfnamefont {B.}~\bibnamefont {Dietz}}, \bibinfo {author} {\bibfnamefont
  {M.}~\bibnamefont {Miski-Oglu}}, \bibinfo {author} {\bibfnamefont
  {A.}~\bibnamefont {Richter}},\ and\ \bibinfo {author} {\bibfnamefont
  {F.}~\bibnamefont {Sch\"afer}},\ }\bibfield  {title} {\bibinfo {title}
  {Distribution of scattering matrix elements in quantum chaotic scattering},\
  }\href {https://doi.org/10.1103/PhysRevLett.111.030403} {\bibfield  {journal}
  {\bibinfo  {journal} {Phys. Rev. Lett.}\ }\textbf {\bibinfo {volume} {111}},\
  \bibinfo {pages} {030403} (\bibinfo {year} {2013})}\BibitemShut {NoStop}%
\bibitem [{\citenamefont {Kuhl}\ \emph {et~al.}(2013)\citenamefont {Kuhl},
  \citenamefont {Legrand},\ and\ \citenamefont {Mortessagne}}]{kuhl13}%
  \BibitemOpen
  \bibfield  {author} {\bibinfo {author} {\bibfnamefont {U.}~\bibnamefont
  {Kuhl}}, \bibinfo {author} {\bibfnamefont {O.}~\bibnamefont {Legrand}},\ and\
  \bibinfo {author} {\bibfnamefont {F.}~\bibnamefont {Mortessagne}},\
  }\bibfield  {title} {\bibinfo {title} {Microwave experiments using open
  chaotic cavities in the realm of the effective Hamiltonian formalism},\
  }\href {https://doi.org/10.1002/prop.201200101} {\bibfield  {journal}
  {\bibinfo  {journal} {Fortschr. Phys.}\ }\textbf {\bibinfo {volume} {61}},\
  \bibinfo {pages} {404} (\bibinfo {year} {2013})}\BibitemShut {NoStop}%
\bibitem [{\citenamefont {Kuhl}\ \emph
  {et~al.}(2005{\natexlab{b}})\citenamefont {Kuhl}, \citenamefont
  {Mart\'{i}nez-Mares}, \citenamefont {M\'{e}ndez-S\'{a}nchez},\ and\
  \citenamefont {St{\"{o}}ckmann}}]{kuhl05}%
  \BibitemOpen
  \bibfield  {author} {\bibinfo {author} {\bibfnamefont {U.}~\bibnamefont
  {Kuhl}}, \bibinfo {author} {\bibfnamefont {M.}~\bibnamefont
  {Mart\'{i}nez-Mares}}, \bibinfo {author} {\bibfnamefont {R.~A.}\ \bibnamefont
  {M\'{e}ndez-S\'{a}nchez}},\ and\ \bibinfo {author} {\bibfnamefont {H.-J.}\
  \bibnamefont {St{\"{o}}ckmann}},\ }\bibfield  {title} {\bibinfo {title}
  {Direct processes in chaotic microwave cavities in the presence of
  absorption},\ }\href {https://doi.org/10.1103/PhysRevLett.94.144101}
  {\bibfield  {journal} {\bibinfo  {journal} {Phys. Rev. Lett.}\ }\textbf
  {\bibinfo {volume} {94}},\ \bibinfo {pages} {144101} (\bibinfo {year}
  {2005}{\natexlab{b}})}\BibitemShut {NoStop}%
\bibitem [{\citenamefont {K\"ober}\ \emph {et~al.}(2010)\citenamefont
  {K\"ober}, \citenamefont {Kuhl}, \citenamefont {St\"ockmann}, \citenamefont
  {Gorin}, \citenamefont {Savin},\ and\ \citenamefont {Seligman}}]{koeb10}%
  \BibitemOpen
  \bibfield  {author} {\bibinfo {author} {\bibfnamefont {B.}~\bibnamefont
  {K\"ober}}, \bibinfo {author} {\bibfnamefont {U.}~\bibnamefont {Kuhl}},
  \bibinfo {author} {\bibfnamefont {H.-J.}\ \bibnamefont {St\"ockmann}},
  \bibinfo {author} {\bibfnamefont {T.}~\bibnamefont {Gorin}}, \bibinfo
  {author} {\bibfnamefont {D.~V.}\ \bibnamefont {Savin}},\ and\ \bibinfo
  {author} {\bibfnamefont {T.~H.}\ \bibnamefont {Seligman}},\ }\bibfield
  {title} {\bibinfo {title} {Microwave fidelity studies by varying antenna
  coupling},\ }\href {https://doi.org/10.1103/PhysRevE.82.036207} {\bibfield
  {journal} {\bibinfo  {journal} {Phys. Rev. E}\ }\textbf {\bibinfo {volume}
  {82}},\ \bibinfo {pages} {036207} (\bibinfo {year} {2010})}\BibitemShut
  {NoStop}%
\bibitem [{\citenamefont {Dietz}\ \emph {et~al.}(2010)\citenamefont {Dietz},
  \citenamefont {Friedrich}, \citenamefont {Harney}, \citenamefont
  {Miski-Oglu}, \citenamefont {Richter}, \citenamefont {Sch\"{a}fer},\ and\
  \citenamefont {Weidenm\"{u}ller}}]{diet10}%
  \BibitemOpen
  \bibfield  {author} {\bibinfo {author} {\bibfnamefont {B.}~\bibnamefont
  {Dietz}}, \bibinfo {author} {\bibfnamefont {T.}~\bibnamefont {Friedrich}},
  \bibinfo {author} {\bibfnamefont {H.~L.}\ \bibnamefont {Harney}}, \bibinfo
  {author} {\bibfnamefont {M.}~\bibnamefont {Miski-Oglu}}, \bibinfo {author}
  {\bibfnamefont {A.}~\bibnamefont {Richter}}, \bibinfo {author} {\bibfnamefont
  {F.}~\bibnamefont {Sch\"{a}fer}},\ and\ \bibinfo {author} {\bibfnamefont
  {H.~A.}\ \bibnamefont {Weidenm\"{u}ller}},\ }\bibfield  {title} {\bibinfo
  {title} {Quantum chaotic scattering in microwave resonators},\ }\href
  {https://doi.org/10.1103/PhysRevE.81.036205} {\bibfield  {journal} {\bibinfo
  {journal} {Phys. Rev. E}\ }\textbf {\bibinfo {volume} {81}},\ \bibinfo
  {pages} {036205} (\bibinfo {year} {2010})}\BibitemShut {NoStop}%
\bibitem [{\citenamefont {Hemmady}\ \emph {et~al.}(2005)\citenamefont
  {Hemmady}, \citenamefont {Zheng}, \citenamefont {Ott}, \citenamefont
  {Antonsen},\ and\ \citenamefont {Anlage}}]{hemm05}%
  \BibitemOpen
  \bibfield  {author} {\bibinfo {author} {\bibfnamefont {S.}~\bibnamefont
  {Hemmady}}, \bibinfo {author} {\bibfnamefont {X.}~\bibnamefont {Zheng}},
  \bibinfo {author} {\bibfnamefont {E.}~\bibnamefont {Ott}}, \bibinfo {author}
  {\bibfnamefont {T.~M.}\ \bibnamefont {Antonsen}},\ and\ \bibinfo {author}
  {\bibfnamefont {S.~M.}\ \bibnamefont {Anlage}},\ }\bibfield  {title}
  {\bibinfo {title} {Universal impedance fluctuations in wave chaotic
  systems},\ }\href {https://doi.org/10.1103/PhysRevLett.94.014102} {\bibfield
  {journal} {\bibinfo  {journal} {Phys. Rev. Lett.}\ }\textbf {\bibinfo
  {volume} {94}},\ \bibinfo {pages} {014102} (\bibinfo {year}
  {2005})}\BibitemShut {NoStop}%
\bibitem [{\citenamefont {Hemmady}\ \emph {et~al.}(2006)\citenamefont
  {Hemmady}, \citenamefont {Zheng}, \citenamefont {Hart}, \citenamefont
  {Antonsen}, \citenamefont {Ott},\ and\ \citenamefont {Anlage}}]{hemm06}%
  \BibitemOpen
  \bibfield  {author} {\bibinfo {author} {\bibfnamefont {S.}~\bibnamefont
  {Hemmady}}, \bibinfo {author} {\bibfnamefont {X.}~\bibnamefont {Zheng}},
  \bibinfo {author} {\bibfnamefont {J.}~\bibnamefont {Hart}}, \bibinfo {author}
  {\bibfnamefont {T.~M.}\ \bibnamefont {Antonsen}}, \bibinfo {author}
  {\bibfnamefont {E.}~\bibnamefont {Ott}},\ and\ \bibinfo {author}
  {\bibfnamefont {S.~M.}\ \bibnamefont {Anlage}},\ }\bibfield  {title}
  {\bibinfo {title} {Universal properties of two-port scattering, impedance,
  and admittance matrices of wave-chaotic systems},\ }\href
  {https://doi.org/10.1103/PhysRevE.74.036213} {\bibfield  {journal} {\bibinfo
  {journal} {Phys. Rev. E}\ }\textbf {\bibinfo {volume} {74}},\ \bibinfo
  {pages} {036213} (\bibinfo {year} {2006})}\BibitemShut {NoStop}%
\bibitem [{\citenamefont {Gradoni}\ \emph {et~al.}(2014)\citenamefont
  {Gradoni}, \citenamefont {Yeh}, \citenamefont {Xiao}, \citenamefont
  {Antonsen}, \citenamefont {Anlage},\ and\ \citenamefont {Ott}}]{grad14}%
  \BibitemOpen
  \bibfield  {author} {\bibinfo {author} {\bibfnamefont {G.}~\bibnamefont
  {Gradoni}}, \bibinfo {author} {\bibfnamefont {J.-H.}\ \bibnamefont {Yeh}},
  \bibinfo {author} {\bibfnamefont {B.}~\bibnamefont {Xiao}}, \bibinfo {author}
  {\bibfnamefont {T.~M.}\ \bibnamefont {Antonsen}}, \bibinfo {author}
  {\bibfnamefont {S.~M.}\ \bibnamefont {Anlage}},\ and\ \bibinfo {author}
  {\bibfnamefont {E.}~\bibnamefont {Ott}},\ }\bibfield  {title} {\bibinfo
  {title} {Predicting the statistics of wave transport through chaotic cavities
  by the random coupling model: A review and recent progress},\ }\href
  {https://doi.org/http://dx.doi.org/10.1016/j.wavemoti.2014.02.003} {\bibfield
   {journal} {\bibinfo  {journal} {Wave Motion}\ }\textbf {\bibinfo {volume}
  {51}},\ \bibinfo {pages} {606} (\bibinfo {year} {2014})}\BibitemShut
  {NoStop}%
\bibitem [{\citenamefont {Kuhl}\ \emph {et~al.}(2008)\citenamefont {Kuhl},
  \citenamefont {H{\"{o}}hmann}, \citenamefont {Main},\ and\ \citenamefont
  {St{\"{o}}ckmann}}]{kuhl08}%
  \BibitemOpen
  \bibfield  {author} {\bibinfo {author} {\bibfnamefont {U.}~\bibnamefont
  {Kuhl}}, \bibinfo {author} {\bibfnamefont {R.}~\bibnamefont {H{\"{o}}hmann}},
  \bibinfo {author} {\bibfnamefont {J.}~\bibnamefont {Main}},\ and\ \bibinfo
  {author} {\bibfnamefont {H.-J.}\ \bibnamefont {St{\"{o}}ckmann}},\ }\bibfield
   {title} {\bibinfo {title} {Resonance widths in open microwave cavities
  studied by harmonic inversion},\ }\href
  {https://doi.org/10.1103/PhysRevLett.100.254101} {\bibfield  {journal}
  {\bibinfo  {journal} {Phys. Rev. Lett.}\ }\textbf {\bibinfo {volume} {100}},\
  \bibinfo {pages} {254101} (\bibinfo {year} {2008})}\BibitemShut {NoStop}%
\bibitem [{\citenamefont {Mahaux}\ and\ \citenamefont
  {Weidenm{\"u}ller}(1969)}]{Mahaux}%
  \BibitemOpen
  \bibfield  {author} {\bibinfo {author} {\bibfnamefont {C.}~\bibnamefont
  {Mahaux}}\ and\ \bibinfo {author} {\bibfnamefont {H.~A.}\ \bibnamefont
  {Weidenm{\"u}ller}},\ }\href@noop {} {\emph {\bibinfo {title} {Shell-model
  Approach to Nuclear Reactions}}}\ (\bibinfo  {publisher} {North-Holland},\
  \bibinfo {address} {Amsterdam},\ \bibinfo {year} {1969})\BibitemShut
  {NoStop}%
\bibitem [{\citenamefont {Verbaarschot}\ \emph {et~al.}(1985)\citenamefont
  {Verbaarschot}, \citenamefont {Weidenm{\"{u}}ller},\ and\ \citenamefont
  {Zirnbauer}}]{verb85}%
  \BibitemOpen
  \bibfield  {author} {\bibinfo {author} {\bibfnamefont {J.~J.~M.}\
  \bibnamefont {Verbaarschot}}, \bibinfo {author} {\bibfnamefont {H.~A.}\
  \bibnamefont {Weidenm{\"{u}}ller}},\ and\ \bibinfo {author} {\bibfnamefont
  {M.~R.}\ \bibnamefont {Zirnbauer}},\ }\bibfield  {title} {\bibinfo {title}
  {Grassmann integration in stochastical quantum physics: The case of
  compound-nucleus scattering},\ }\href
  {https://doi.org/10.1016/0370-1573(85)90070-5} {\bibfield  {journal}
  {\bibinfo  {journal} {Phys. Rep.}\ }\textbf {\bibinfo {volume} {129}},\
  \bibinfo {pages} {367} (\bibinfo {year} {1985})}\BibitemShut {NoStop}%
\bibitem [{\citenamefont {Sokolov}\ and\ \citenamefont
  {Zelevinsky}(1989)}]{soko89}%
  \BibitemOpen
  \bibfield  {author} {\bibinfo {author} {\bibfnamefont {V.~V.}\ \bibnamefont
  {Sokolov}}\ and\ \bibinfo {author} {\bibfnamefont {V.~G.}\ \bibnamefont
  {Zelevinsky}},\ }\bibfield  {title} {\bibinfo {title} {Dynamics and
  statistics of unstable quantum states},\ }\href
  {https://doi.org/10.1016/0375-9474(89)90558-7} {\bibfield  {journal}
  {\bibinfo  {journal} {Nucl. Phys. A}\ }\textbf {\bibinfo {volume} {504}},\
  \bibinfo {pages} {562} (\bibinfo {year} {1989})}\BibitemShut {NoStop}%
\bibitem [{\citenamefont {Fyodorov}\ and\ \citenamefont
  {Sommers}(1997)}]{fyod97}%
  \BibitemOpen
  \bibfield  {author} {\bibinfo {author} {\bibfnamefont {Y.~V.}\ \bibnamefont
  {Fyodorov}}\ and\ \bibinfo {author} {\bibfnamefont {H.-J.}\ \bibnamefont
  {Sommers}},\ }\bibfield  {title} {\bibinfo {title} {Statistics of resonance
  poles, phase shifts and time delays in quantum chaotic scattering: Random
  matrix approach for systems with broken time-reversal invariance},\ }\href
  {https://doi.org/10.1063/1.531919} {\bibfield  {journal} {\bibinfo  {journal}
  {J. Math. Phys.}\ }\textbf {\bibinfo {volume} {38}},\ \bibinfo {pages} {1918}
  (\bibinfo {year} {1997})}\BibitemShut {NoStop}%
\bibitem [{\citenamefont {St{\"{o}}ckmann}(1999)}]{Stoeckmann}%
  \BibitemOpen
  \bibfield  {author} {\bibinfo {author} {\bibfnamefont {H.-J.}\ \bibnamefont
  {St{\"{o}}ckmann}},\ }\href@noop {} {\emph {\bibinfo {title} {Quantum Chaos:
  An Introduction}}}\ (\bibinfo  {publisher} {Cambridge University Press},\
  \bibinfo {address} {Cambridge, UK},\ \bibinfo {year} {1999})\BibitemShut
  {NoStop}%
\bibitem [{\citenamefont {Guhr}\ \emph {et~al.}(1998)\citenamefont {Guhr},
  \citenamefont {{M\"{u}ller-Groeling}},\ and\ \citenamefont
  {Weidenm{\"{u}}ller}}]{guhr98}%
  \BibitemOpen
  \bibfield  {author} {\bibinfo {author} {\bibfnamefont {T.}~\bibnamefont
  {Guhr}}, \bibinfo {author} {\bibfnamefont {A.}~\bibnamefont
  {{M\"{u}ller-Groeling}}},\ and\ \bibinfo {author} {\bibfnamefont {H.~A.}\
  \bibnamefont {Weidenm{\"{u}}ller}},\ }\bibfield  {title} {\bibinfo {title}
  {Random matrix theories in {Q}uantum {P}hysics: Common concepts},\ }\href
  {https://doi.org/10.1016/S0370-1573(97)00088-4} {\bibfield  {journal}
  {\bibinfo  {journal} {Phys. Rep.}\ }\textbf {\bibinfo {volume} {299}},\
  \bibinfo {pages} {189} (\bibinfo {year} {1998})}\BibitemShut {NoStop}%
\bibitem [{\citenamefont {Mitchell}\ \emph {et~al.}(2010)\citenamefont
  {Mitchell}, \citenamefont {Richter},\ and\ \citenamefont
  {Weidenm\"uller}}]{mitc10}%
  \BibitemOpen
  \bibfield  {author} {\bibinfo {author} {\bibfnamefont {G.~E.}\ \bibnamefont
  {Mitchell}}, \bibinfo {author} {\bibfnamefont {A.}~\bibnamefont {Richter}},\
  and\ \bibinfo {author} {\bibfnamefont {H.~A.}\ \bibnamefont
  {Weidenm\"uller}},\ }\bibfield  {title} {\bibinfo {title} {Random matrices
  and chaos in nuclear physics: Nuclear reactions},\ }\href
  {https://doi.org/10.1103/RevModPhys.82.2845} {\bibfield  {journal} {\bibinfo
  {journal} {Rev. Mod. Phys.}\ }\textbf {\bibinfo {volume} {82}},\ \bibinfo
  {pages} {2845} (\bibinfo {year} {2010})}\BibitemShut {NoStop}%
\bibitem [{\citenamefont {Fyodorov}\ and\ \citenamefont
  {Savin}(2011)}]{fyod11ox}%
  \BibitemOpen
  \bibfield  {author} {\bibinfo {author} {\bibfnamefont {Y.~V.}\ \bibnamefont
  {Fyodorov}}\ and\ \bibinfo {author} {\bibfnamefont {D.~V.}\ \bibnamefont
  {Savin}},\ }\bibfield  {title} {\bibinfo {title} {Resonance scattering of
  waves in chaotic systems},\ }in\ \href@noop {} {\emph {\bibinfo {booktitle}
  {The Oxford Handbook of Random Matrix Theory}}},\ \bibinfo {editor} {edited
  by\ \bibinfo {editor} {\bibfnamefont {G.}~\bibnamefont {Akemann}}, \bibinfo
  {editor} {\bibfnamefont {J.}~\bibnamefont {Baik}},\ and\ \bibinfo {editor}
  {\bibfnamefont {P.}~\bibnamefont {Di~Francesco}}}\ (\bibinfo  {publisher}
  {Oxford University Press, UK},\ \bibinfo {year} {2011})\ Chap.~\bibinfo
  {chapter} {34}, pp.\ \bibinfo {pages} {703--722},\ \bibinfo {note}
  {[arXiv:1003.0702]}\BibitemShut {NoStop}%
\bibitem [{\citenamefont {Savin}(2018)}]{savi18}%
  \BibitemOpen
  \bibfield  {author} {\bibinfo {author} {\bibfnamefont {D.~V.}\ \bibnamefont
  {Savin}},\ }\bibfield  {title} {\bibinfo {title} {Envelope and phase
  distribution of a resonance transmission through a complex environment},\
  }\href {https://doi.org/10.1103/PhysRevE.97.062202} {\bibfield  {journal}
  {\bibinfo  {journal} {Phys. Rev. E}\ }\textbf {\bibinfo {volume} {97}},\
  \bibinfo {pages} {062202} (\bibinfo {year} {2018})}\BibitemShut {NoStop}%
\bibitem [{\citenamefont {Fyodorov}\ and\ \citenamefont
  {Savin}(2004)}]{fyod04b}%
  \BibitemOpen
  \bibfield  {author} {\bibinfo {author} {\bibfnamefont {Y.~V.}\ \bibnamefont
  {Fyodorov}}\ and\ \bibinfo {author} {\bibfnamefont {D.~V.}\ \bibnamefont
  {Savin}},\ }\bibfield  {title} {\bibinfo {title} {Statistics of impedance,
  local density of states, and reflection in quantum chaotic systems with
  absorption},\ }\href {https://doi.org/10.1134/1.1868794} {\bibfield
  {journal} {\bibinfo  {journal} {JETP Lett.}\ }\textbf {\bibinfo {volume}
  {80}},\ \bibinfo {pages} {725} (\bibinfo {year} {2004})}\BibitemShut
  {NoStop}%
\bibitem []{note1}%
  \BibitemOpen
  {The background acts as a
  single-channel scattering centre, with the reflection amplitude $S_\protect
  \mathrm {bg}=\protect \frac {1-i\eta K}{1+i\eta K}$. At $\eta =1$, the
  reflection coefficient $|S_\protect \mathrm {bg}|^2=\protect \frac
  {x-1}{x+1}<1$ becomes statistically independent of the reflection phase,
  yielding (\ref {P(u,v)}), see \cite{fyod04b}. Note that $S_\protect \mathrm
  {bg}$ is subunitary at finite absorption, resulting in subunitary
  $S$.}\BibitemShut {Stop}%
\bibitem [{\citenamefont {Savin}\ \emph {et~al.}(2005)\citenamefont {Savin},
  \citenamefont {Sommers},\ and\ \citenamefont {Fyodorov}}]{savi05}%
  \BibitemOpen
  \bibfield  {author} {\bibinfo {author} {\bibfnamefont {D.~V.}\ \bibnamefont
  {Savin}}, \bibinfo {author} {\bibfnamefont {H.-J.}\ \bibnamefont {Sommers}},\
  and\ \bibinfo {author} {\bibfnamefont {Y.~V.}\ \bibnamefont {Fyodorov}},\
  }\bibfield  {title} {\bibinfo {title} {Universal statistics of the local
  Green's function in wave chaotic systems with absorption},\ }\href
  {https://doi.org/10.1134/1.2150877} {\bibfield  {journal} {\bibinfo
  {journal} {JETP Lett.}\ }\textbf {\bibinfo {volume} {82}},\ \bibinfo {pages}
  {544} (\bibinfo {year} {2005})}\BibitemShut {NoStop}%
\bibitem []{note2}%
  \BibitemOpen
  {Choosing $p=(r-r_{-})/(1-r-t)\ge0$ as a new integration variable, one can further
  bring (\ref{Ptr}) to the form of Ref.~\cite{savi17}.}\BibitemShut {Stop}%
\bibitem []{note3}%
  \BibitemOpen
  {It is worth noting a matrix relation $1-S^\dagger S = (1-S^{(0)})d$.}
  \BibitemShut {Stop}%
\bibitem [{\citenamefont {Beenakker}\ and\ \citenamefont
  {Brouwer}(2001)}]{been01}%
  \BibitemOpen
  \bibfield  {author} {\bibinfo {author} {\bibfnamefont {C.~W.~J.}\
  \bibnamefont {Beenakker}}\ and\ \bibinfo {author} {\bibfnamefont {P.~W.}\
  \bibnamefont {Brouwer}},\ }\bibfield  {title} {\bibinfo {title} {Distribution
  of the reflection eigenvalues of a weakly absorbing chaotic cavity},\
  }\href {https://doi.org/10.1016/S1386-9477(00)00245-9}
  {\bibfield  {journal} {\bibinfo  {journal} {Physica E}\
  }\textbf {\bibinfo {volume} {9}},\ \bibinfo {pages} {463} (\bibinfo {year}
  {2001})}\BibitemShut {NoStop}%
\bibitem [{\citenamefont {Mirlin}\ \emph {et~al.}(2006)\citenamefont {Mirlin},
  \citenamefont {Fyodorov}, \citenamefont {Mildenberger},\ and\ \citenamefont
  {Evers}}]{mirl06}%
  \BibitemOpen
  \bibfield  {author} {\bibinfo {author} {\bibfnamefont {A.~D.}\ \bibnamefont
  {Mirlin}}, \bibinfo {author} {\bibfnamefont {Y.~V.}\ \bibnamefont
  {Fyodorov}}, \bibinfo {author} {\bibfnamefont {A.}~\bibnamefont
  {Mildenberger}},\ and\ \bibinfo {author} {\bibfnamefont {F.}~\bibnamefont
  {Evers}},\ }\bibfield  {title} {\bibinfo {title} {Exact relations between
  multifractal exponents at the {A}nderson transition},\ }\href
  {https://doi.org/10.1103/PhysRevLett.97.046803} {\bibfield  {journal}
  {\bibinfo  {journal} {Phys. Rev. Lett.}\ }\textbf {\bibinfo {volume} {97}},\
  \bibinfo {pages} {046803} (\bibinfo {year} {2006})}\BibitemShut {NoStop}%
\bibitem [{\citenamefont {Gruzberg}\ \emph {et~al.}(2011)\citenamefont
  {Gruzberg}, \citenamefont {Ludwig}, \citenamefont {Mirlin},\ and\
  \citenamefont {Zirnbauer}}]{gruz11}%
  \BibitemOpen
  \bibfield  {author} {\bibinfo {author} {\bibfnamefont {I.~A.}\ \bibnamefont
  {Gruzberg}}, \bibinfo {author} {\bibfnamefont {A.~W.~W.}\ \bibnamefont
  {Ludwig}}, \bibinfo {author} {\bibfnamefont {A.~D.}\ \bibnamefont {Mirlin}},\
  and\ \bibinfo {author} {\bibfnamefont {M.~R.}\ \bibnamefont {Zirnbauer}},\
  }\bibfield  {title} {\bibinfo {title} {Symmetries of multifractal spectra and
  field theories of {A}nderson localization},\ }\href
  {https://doi.org/10.1103/PhysRevLett.107.086403} {\bibfield  {journal}
  {\bibinfo  {journal} {Phys. Rev. Lett.}\ }\textbf {\bibinfo {volume} {107}},\
  \bibinfo {pages} {086403} (\bibinfo {year} {2011})}\BibitemShut {NoStop}%
\bibitem [{\citenamefont {Khaymovich}\ \emph {et~al.}(2015)\citenamefont
  {Khaymovich}, \citenamefont {Koski}, \citenamefont {Saira}, \citenamefont
  {Kravtsov},\ and\ \citenamefont {Pekola}}]{khay15}%
  \BibitemOpen
  \bibfield  {author} {\bibinfo {author} {\bibfnamefont {I.}~\bibnamefont
  {Khaymovich}}, \bibinfo {author} {\bibfnamefont {J.}~\bibnamefont {Koski}},
  \bibinfo {author} {\bibfnamefont {O.-P.}\ \bibnamefont {Saira}}, \bibinfo
  {author} {\bibfnamefont {V.}~\bibnamefont {Kravtsov}},\ and\ \bibinfo
  {author} {\bibfnamefont {J.}~\bibnamefont {Pekola}},\ }\bibfield  {title}
  {\bibinfo {title} {Multifractality of random eigenfunctions and
  generalization of Jarzynski equality},\ }\href
  {https://doi.org/10.1038/ncomms8010} {\bibfield  {journal} {\bibinfo
  {journal} {Nat. Comms.}\ }\textbf {\bibinfo {volume} {6}},\ \bibinfo {pages}
  {7010} (\bibinfo {year} {2015})}\BibitemShut {NoStop}%
\bibitem [{\citenamefont {Savin}\ \emph {et~al.}(2006)\citenamefont {Savin},
  \citenamefont {Legrand},\ and\ \citenamefont {Mortessagne}}]{savi06b}%
  \BibitemOpen
  \bibfield  {author} {\bibinfo {author} {\bibfnamefont {D.~V.}\ \bibnamefont
  {Savin}}, \bibinfo {author} {\bibfnamefont {O.}~\bibnamefont {Legrand}},\
  and\ \bibinfo {author} {\bibfnamefont {F.}~\bibnamefont {Mortessagne}},\
  }\bibfield  {title} {\bibinfo {title} {Inhomogeneous losses and complexness
  of wave functions in chaotic cavities},\ }\href
  {https://doi.org/10.1209/epl/i2006-10358-3} {\bibfield  {journal} {\bibinfo
  {journal} {Europhys. Lett.}\ }\textbf {\bibinfo {volume} {76}},\ \bibinfo
  {pages} {774} (\bibinfo {year} {2006})}\BibitemShut {NoStop}%
\bibitem [{\citenamefont {Savin}(2017{\natexlab{b}})}]{savi17appa}%
  \BibitemOpen
  \bibfield  {author} {\bibinfo {author} {\bibfnamefont {D.~V.}\ \bibnamefont
  {Savin}},\ }\bibfield  {title} {\enquote {\bibinfo {title} {Fluctuations and
  correlations in scattering on a resonance coupled to a chaotic background},}\
  }\href {\doibase 10.12693/APhysPolA.132.1688} {\bibfield  {journal} {\bibinfo
   {journal} {Acta Phys. Pol. A}\ }\textbf {\bibinfo {volume} {132}},\ \bibinfo
  {pages} {1688} (\bibinfo {year} {2017}{\natexlab{b}})}\BibitemShut {NoStop}%
\bibitem [{\citenamefont {Lawniczak}\ and\ \citenamefont
  {Sirko}(2019)}]{lawn19}%
  \BibitemOpen
  \bibfield  {author} {\bibinfo {author} {\bibfnamefont {M.}~\bibnamefont
  {Lawniczak}}\ and\ \bibinfo {author} {\bibfnamefont {L.}~\bibnamefont
  {Sirko}},\ }\bibfield  {title} {\bibinfo {title} {Investigation of the
  diagonal elements of the {Wigner's} reaction matrix for networks with
  violated time reversal invariance},\ }\href
  {https://doi.org/10.1038/s41598-019-42123-y} {\bibfield  {journal} {\bibinfo
  {journal} {Sci. Rep.}\ }\textbf {\bibinfo {volume} {9}},\ \bibinfo {pages}
  {5630} (\bibinfo {year} {2019})}\BibitemShut {NoStop}%
\end{thebibliography}
%\end{document}

\end{document}